\documentclass[aps,prb,reprint,nofootinbib,amsmath,amssymb,superscriptaddress]{revtex4-2}

\usepackage{amsmath,amssymb,amsthm,bm,physics,mathtools,graphicx,booktabs,enumitem,hyperref,float}

\usepackage[margin=0.75in,columnsep=0.3in]{geometry}

\newcommand{\avg}[1]{\langle #1\rangle}
\begin{document}

\title{Dynamic Pseudogap Model: Analytic Results vs  Monte Carlo Numerics}

\author{Michael V. Sadovskii}

\affiliation{Institute for Electrophysics, Russian Academy of Sciences,
Ural Branch,\\
Amundsen str. 106, Ekaterinburg 620016, Russia\\
E-mail: sadovski@iep.uran.ru}


\begin{abstract}

\noindent

Recently proposed dynamic pseudogap model analytical results
for spectral density of an electron scattered by random time-dependent Gaussian
field representing fluctuations with finite nesting vector ${\bf Q}$,
characteristic oscillation frequency $\omega_0$ and damping $\gamma$,
leading to pseudogap formation, are compared with exact numerical solution
by Monte Carlo (MC) approach. We consider both commensurate (real) and
incommensurate (complex) fluctuations, for a wide range of parameters of the
model. Electron Green's function is approximated either by a double Poisson
series with history-dependent Bogoliubov gaps or by exact static continued
fraction solution. In the coherent regime double-series representation provides
description of dynamic sideband formation in the spectral density.
As coherence is progressively lost, the theory crosses over to the quasistatic
fluctuating-gap regime described by the exact continued-fraction solution.
The exact MC solution overcomes the limitations of analytic approach, such as
neglect of Magnus-type corrections in commuting approximation, 
and establish the limits of analytic approach, as well as confirming the 
exactness of continued fraction solution in static limit. 
In general, analytic expressions demonstrate rather satisfactory agreement with 
exact MC numerics within their limits, confirming the overall qualitative 
physical picture of dynamic pseudogap formation.

\end{abstract}

\maketitle


\section{Introduction}

Pseudogap formation in metals with strong short-range order fluctuations
(charge or spin density waves, excitonic and Peierls fluctuations, underdoped
cuprates) has a class of exactly solvable one-dimensional models in which
electrons scatter from a static Gaussian random field with finite nesting vector
$\mathbf Q$ and finite correlation length~\cite{Sadovskii1974,Sadovskii1979,
Sadovskii2019}. Generalizing these to \emph{dynamic} fluctuations is an open
problem, while the characteristic fluctuation frequency may become 
comparable to electronic energies near structural or electronic transitions.
For a spatially uniform field ($Q=0$) the problem is exactly
solvable~\cite{KS2024,KiselevKikoin2009}.
For a finite nesting vector $\mathbf Q$, every scattering event alternates the
electron between two branches of electron spectrum  $\xi$ and $-\xi$,
producing a non-Abelian SU(2) pseudospin structure and Hamiltonians that do not
commute at different times.

Recently an analytical description of this model was proposed~\cite{KS2026}
neglecting the non-commutativity (the ordering of pseudospin vertices),
summing the Gaussian averages into complex Poisson weights and associating each
stochastic history  with an effective complex gap $\Delta_{nm}$ in a
Bogoliubov--Gor'kov propagator. The result is a coherent superposition of
dynamically broadened Bogoliubov branches (the ``double series''), with the
decoherence rate $\Gamma_{\rm eff}=\gamma\Delta^2/(\omega_0^2+\gamma^2)$
(where $\Delta$ is random field amplitude), controlling a crossover to the
static continued-fraction solution. The construction is explicitly
heuristic and is supported in~\cite{KS2026} by its exact limits and by
comparison with the static ($\omega_0=0$) solution.

Here we test these analytic results against a numerically exact solution of
the same model. 
The averaged electron propagator is computed to controlled
statistical accuracy. This allows a direct test of the validity range and
limitations of analytic theory and of the crossover picture away from
the exact limits.

\section{Model and method}

\label{sec:model}

The Hamiltonian in Nambu space is
\begin{equation}
\hat H(t)=\xi\sigma_3+\Delta_1(t)\sigma_1+\Delta_2(t)\sigma_2,
\label{Hamilt}
\end{equation}
where dynamic random field $\Delta=\Delta_1+i\Delta_2$
is a zero-mean Gaussian field described by correlator
\begin{equation}
\avg{\Delta(t)\Delta^*(0)}=\Delta^2e^{-\gamma|t|}\cos\omega_0t.
\label{eq:corr}
\end{equation}
Commensurate order corresponds to real field, so that $\Delta_2=0$, while
incommensurate is described by complex two-component $\Delta_1+i\Delta_2$
field.

The general form of retarded retarded Green's function proposed in
Ref. \cite{KS2026} has the Bogoliubov -- Gor'kov form, represented by
the following double series:
\begin{widetext}
\begin{equation}
G^R(\omega,\xi)
=\sum_{n,m=0}^{\infty}
P_{\lambda_1}(n)P_{\lambda_2}(m)
\frac{u_{nm}^2}
{\omega+i\Gamma_{\rm eff}-\sqrt{\xi^2+\Delta_{nm}^2}}
+\sum_{n,m=0}^{\infty}
P_{\lambda_1}^*(n)P_{\lambda_2}^*(m)
\frac{(v^{2}_{nm})^*}
{\omega+i\Gamma_{\rm eff}+\sqrt{\xi^2+(\Delta^{2}_{nm})^*}}
\label{BogolFreqDoubleSer}
\end{equation}
\end{widetext}
where Bogoliubov coefficients are:
\begin{equation}
u_{nm}^2
=
\frac12
\left(
1+\frac{\xi}{E_{nm}}
\right),
\end{equation}
and
\begin{equation}
v_{nm}^2
=
\frac12
\left(
1-\frac{\xi}{E_{nm}}
\right).
\end{equation}
which satisfy
\begin{equation}
u_{nm}^2+v_{nm}^2=1.
\end{equation}
The complex quasiparticle energy is defined as
\begin{equation}
E_{nm}
=
\sqrt{\xi^2+\Delta_{nm}^2},
\label{eq:EnmDS}
\end{equation}
Since $E_{nm}$ is complex, Bogoliubov $u-v$ become complex amplitudes.
They therefore cannot be interpreted as occupation probabilities.
Instead they determine the relative weights of electron-like and hole-like
components within each coherent scattering history defined by $(n,m)$.
The complete Green's function is obtained by coherent summation over all
histories.

Two dynamic matrix scattering fields in Nambu space are defined as:
\begin{eqnarray}
\hat\Delta_{1n}=\sigma_1\Delta_{1n}=\sigma_1(\omega_0-i\gamma)(n-\lambda_1),
\nonumber\\
\hat\Delta_{2m}=\sigma_1\Delta_{2m}=\sigma_1(-\omega_0-i\gamma)(m-\lambda_2),
\label{eq:delta12DS}
\end{eqnarray}
and are distributed according to complex Poisson distributions:
\begin{equation}
P_{\lambda_i}(k)=\frac{\lambda_i^k}{k!}\exp(-\lambda_i),\, i=1,2
\label{Poiss}
\end{equation}
where
\begin{equation}
\lambda_1
=
\frac{\Delta^2}
{2(\omega_0-i\gamma)^2},
\qquad
\lambda_2
=
\frac{\Delta^2}
{2(-\omega_0-i\gamma)^2}.
\label{lambi}
\end{equation}
The corresponding complex pseudogap scale is determined by $\Delta_{nm}$, where
for commensurate case
\begin{equation}
\Delta_{nm}
=
\Delta_{1n}+\Delta_{2m}.
\label{eq:LambdaCommDS}
\end{equation}
and the acting Poissonian field is
\begin{equation}
\hat\Delta_{nm}=\sigma_1\Delta_{nm}
=
\sigma_1(\omega_0-i\gamma)(n-\lambda_1)
+
\sigma_1(-\omega_0-i\gamma)(m-\lambda_2).
\label{DynamicGap}
\end{equation}
Note that for commensurate case
\begin{equation}
\Delta_{nm}=
(n-m)\omega_0
-i(n+m)\gamma - i\Gamma_{\mathrm{eff}}.
\label{Delta00}
\end{equation}
contains the effective dynamic decoherence rate
\begin{equation}
\Gamma_{\mathrm{eff}}
=
\frac{\gamma\Delta^2}
{\omega_0^2+\gamma^2}.
\label{GammaEff}
\end{equation}
which plays a central role in our theory\footnote{In combination with
$\Gamma_{\mathrm{eff}}$ entering the denominator of (\ref{BogolFreqDoubleSer})
this creates double-counting problem, breaking the exact solution at 
the Fermi surface at $\xi=0$, it is solved \cite{KS2026} in numerical 
calculations for commensurate case by appropriate subtraction 
from $\Delta_{nm}$.}.

For incommensurate case $\Delta_{nm}$ is represented by two-dimensional
pseudospin vector in Nambu space:
\begin{equation}
\mathbf{\Delta}_{nm}
=
(\Delta_{1n},\Delta_{2m}),
\end{equation}
whose components are introduced as
\begin{equation}
\Delta_{1n}
=
(\omega_0-i\gamma)
(n-\lambda_1),
\label{Gap1}
\end{equation}
and
\begin{equation}
\Delta_{2m}
=
(-\omega_0-i\gamma)
(m-\lambda_2).
\label{Gap2}
\end{equation}
and effective gap entering the history Green's function defined in
(\ref{BogolFreqDoubleSer} ) must preserve the $O(2)$ symmetry in pseudospin
space:
\begin{equation}
\mathbf{\Delta}_{nm}^2
=
\Delta_{1n}^2
+
\Delta_{2m}^2,
\label{DynamInvariant}
\end{equation}
with two Cartesian components in Nambu space generating two statistically
independent stochastic histories.

Average gap is easily shown to be zero
\begin{eqnarray}
\left<{\Delta_{nm}}\right>=\sum_{n,m=0}^\infty P_{\lambda_1}(n)P_{\lambda_2}(m)
\times\nonumber\\
\times[(\omega_0-i\gamma)(n-\lambda_1)+(-\omega_0-i\gamma)(m-\lambda_2)]=0
\nonumber\\
\label{zero_gap}
\end{eqnarray}
corresponding to the absence of long-range order and vanishing anomalous
Green's functions.

The analysis of Ref. \cite{KS2026} suggested that the dynamic pseudogap problem
possesses the following physically distinct regimes.
\begin{enumerate}
\item
For $\omega_0\gg\Gamma_{\rm eff}$,
the spectrum is governed by coherent dynamic sidebands described by the
double-series representation.
\item
For $\omega_0\sim\Gamma_{\rm eff}$,
the sidebands overlap and coherence is progressively lost, and double series
just can not be applied here. It is clear that calculated spectra should evolve
continuously toward  static ($\omega_0=0$) continued-fraction solution.
\item
For $\gamma\gg\omega_0$, motional narrowing \cite{Kubo} 
suppresses pseudogap
formation altogether and restores conventional free quasiparticle behavior.
\end{enumerate}

The static $\omega_0=0$ theory thus appears naturally as quasistatic limit of
the more general dynamic theory. It is described by summation of the complete
perturbation series \cite{KS2026}, similar to that obtained long ago in
Ref. \cite{Sadovskii1979}, which reduces to the exact recursion relation
for the irreducible self-energy,
\begin{equation}
\Sigma_k(\omega,\xi)
=
\frac{\Delta^2v(k)}
{\omega-(-1)^k\xi
+ik\gamma
-\Sigma_{k+1}(\omega,\xi)},
\label{Si_rec}
\end{equation}
which immediately yields the recursion relation for the Green's function,
\begin{equation}
G_k(\omega,\xi)
=
\left[
\omega
-
(-1)^k\xi
+
ik\gamma
-
\Delta^2v(k+1)
G_{k+1}
(\omega,\xi)
\right]^{-1}.
\label{G_rec}
\end{equation}
where combinatorial factors $v(k)$ count the number of topologically equivalent
diagrams.
For commensurate fluctuations,
\begin{equation}
v(k)=k,
\label{vcomm}
\end{equation}
whereas for incommensurate fluctuations
\begin{equation}
v(k)=
\left\{
\begin{array}{ll}
\dfrac{k+1}{2},
&
k\ {\rm odd},
\\[2mm]
\dfrac{k}{2},
&
k\ {\rm even}.
\end{array}
\right.
\label{vinc}
\end{equation}
The physical Green's function is obtained from
$G(\omega,\xi)=G_0(\omega,\xi)$.
Equivalently, one obtains the exact continued-fraction representation similar
to that of Ref. \cite{Sadovskii1979}:
\small
\begin{equation}
G(\omega,\xi)
=
\frac{1}
{\displaystyle
\omega-\xi-
\frac{v(1)\Delta^2}
{\displaystyle
\omega+\xi+i\gamma-
\frac{v(2)\Delta^2}
{\displaystyle
\omega-\xi+2i\gamma-
\frac{v(3)\Delta^2}
{\displaystyle
\omega+\xi+3i\gamma-\cdots}}}}
\label{G_chain}
\end{equation}
\normalsize
Equations (\ref{Si_rec})--(\ref{G_chain}) provide the exact solution of
the quasistatic dynamic pseudogap model.

To determine precisely the validity and limitations of this theory we have
undertaken extensive Monte Carlo calculations leading to numerically exact
solution of our model.
The averaged retarded Green's function in time representation is given by
\begin{equation}
G^R(t)=-i\avg{U_{11}(t)}
\label{Gt}
\end{equation}
where time-evolution operator
\begin{equation}
\hat U=T\exp[-i\int_0^t\hat H\,dt']
\label{Ut}
\end{equation}
is to be calculated for many realizations of the random field with matrix
Hamiltonian given in (\ref{Hamilt}).

The spectral density was calculated as:
\begin{equation}
A(\omega)=\frac1\pi\,\mathrm{Re}\int_0^Tdt\,e^{i\omega t-\eta t}\avg{U_{11}(t)} ,
\label{spd}
\end{equation}
where $\eta$ is a small broadening applied identically to all methods compared.

The field is generated exactly: the commensurate field is $\mathrm{Re}\,z$
with $z$ a complex Ornstein--Uhlenbeck (OU) process of decay $\gamma+i\omega_0$;
the incommensurate field is the sum of two independent complex OU processes
with decay $\gamma\pm i\omega_0$. The propagator is advanced by the exact
$2\times2$ step exponential with the field averaged over each step, and averaged
over $1.5$--$3\times10^4$ realizations. No commuting approximation is made,
so that Magnus terms \cite{Magnus}, neglected in Ref.\cite{KS2026} are actually 
accounted here. Details, parameters and checks can be found in Supplement S1.
We also made comparison with the results of double series
(\ref{BogolFreqDoubleSer}) calculations of Ref.~\cite{KS2026}, in its retarded
form with conjugated weights, and the results obtained from continued fraction
(\ref{G_chain}) at $\omega_0=0$ \cite{KS2026}] with the same values of $\gamma$.
All energies below are are in units of $\Delta$.

\section{Results}

\subsection{Exact limits and consistency checks}

\begin{figure*}[t]
\centering
\includegraphics[width=0.75\textwidth]{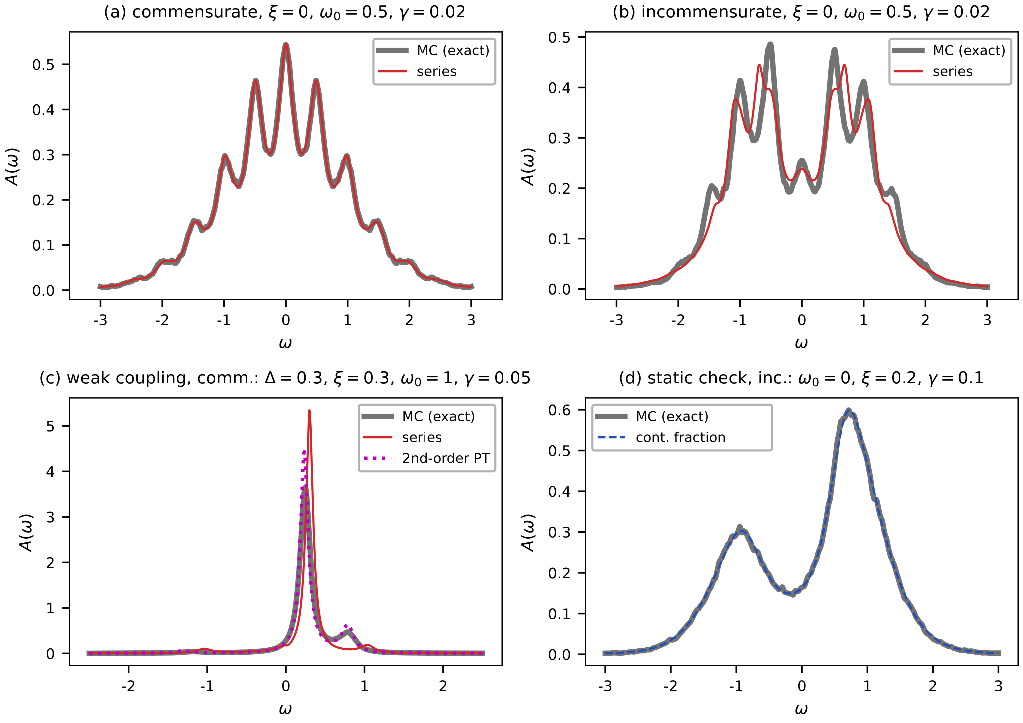}
\caption{Spectral density: consistency checks.
(a) Commensurate case, $\xi=0$, $\omega_0=0.5$, $\gamma=0.02$: the series is exact.
(b) Incommensurate case, same parameters: small deviations from the non-Abelian term.
(c) Weak coupling $\Delta=0.3$, $\xi=0.3$, $\omega_0=1$, $\gamma=0.05$:
MC (grey), series (red), second-order perturbation theory (green dotted).
(d) Static check ($\omega_0=0$, incommensurate, $\xi=0.2$, $\gamma=0.1$):
MC against the continued fraction.}
\label{fig:bench}
\end{figure*}

Figure~\ref{fig:bench} shows four checks. (a) For commensurate fluctuations at
$\xi=0$, where the Hamiltonians commute and the theory is exact, the series
agrees with the MC spectrum to within the noise (maximum deviation $0.013$).
(b) For incommensurate fluctuations at $\xi=0$ the series is not exact, since
the Magnus terms \cite{KS2026} do not vanish; the maximum deviation
is $0.09$--$0.13$ with the sideband weights and positions slightly misplaced.
(d) For the static case $\omega_0=0$ the continued fraction agrees with MC to
$0.015$--$0.027$ (within noise), supporting its exactness for the exponentially
correlated field with the incommensurate vertex factors $v(k)$ of~\cite{KS2026};
the same holds for the commensurate case.

\subsection{Comparison for a wide range of parametetrs.}

Figures~\ref{fig:pp1}--\ref{fig:pp4} below reproduce the parameter sets used in
Ref.~\cite{KS2026} with MC data added. We can see that the overall qualitative
physical picture is the same in all cases.

\begin{widetext}

\begin{figure}[H]\centering\includegraphics[width=0.70\textwidth]
{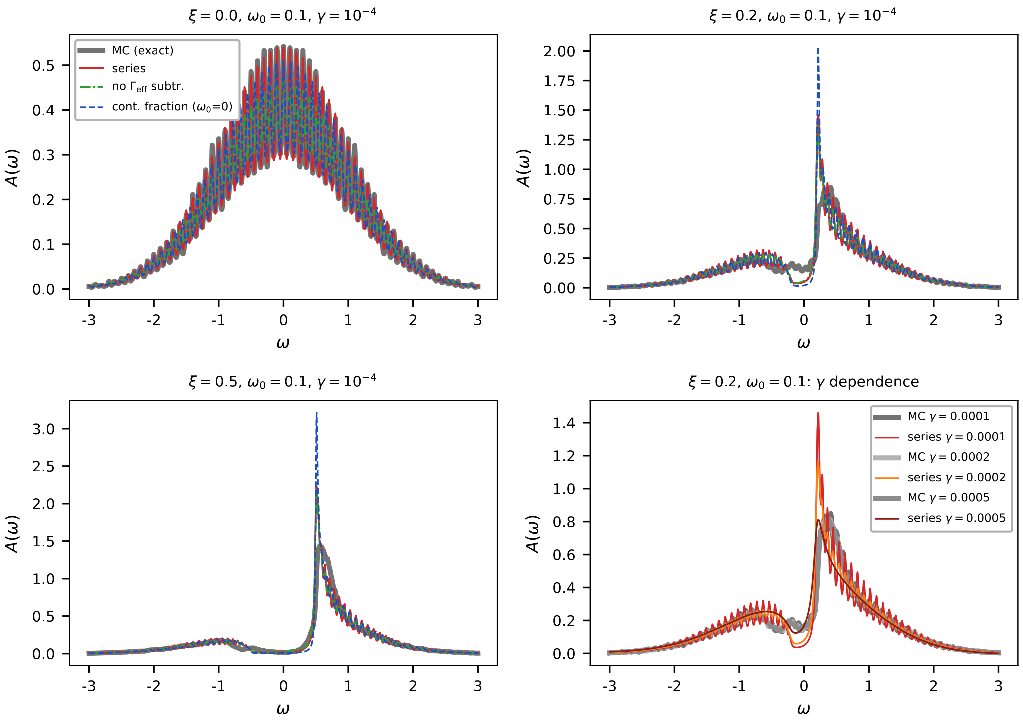}
\caption{Spectral density: commensurate case, 
$\omega_0=0.1$.}
\label{fig:pp1}
\end{figure}

\begin{figure}[H]\centering\includegraphics[width=0.70\textwidth]
{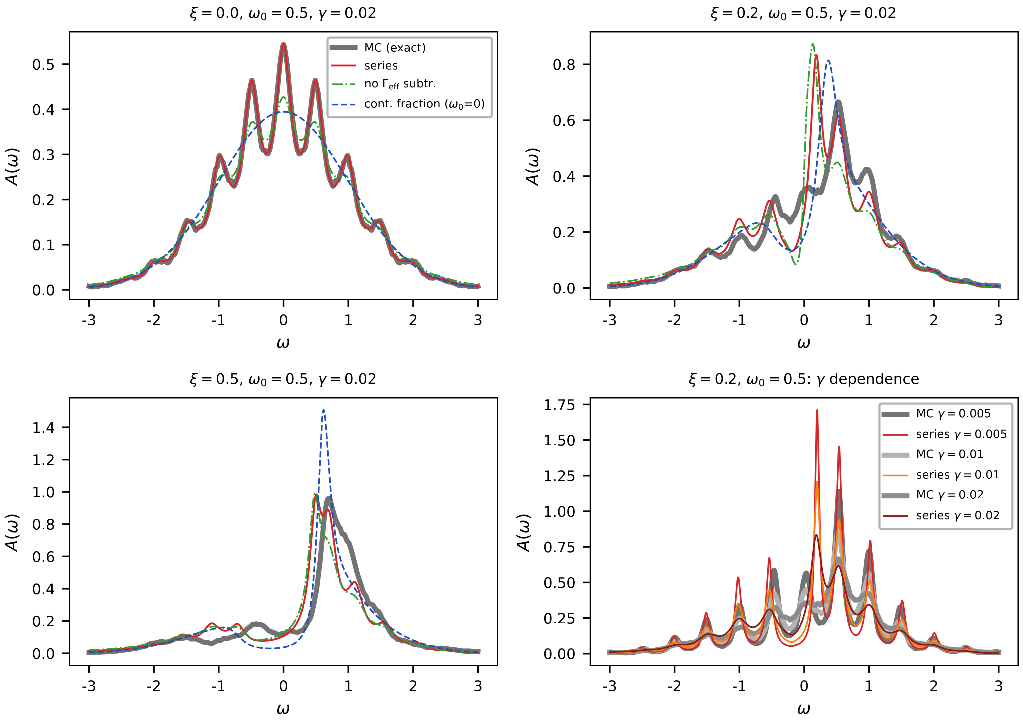}
\caption{Spectral density: commensurate case, $\omega_0=0.5$.
Dashed green lines show here and below the data obtained from double
series with no $\Gamma_{\rm eff}$ subtraction in $\Delta_{nm}$.}
\label{fig:pp2}
\end{figure}

\begin{figure}[H]\centering\includegraphics[width=0.65\textwidth]
{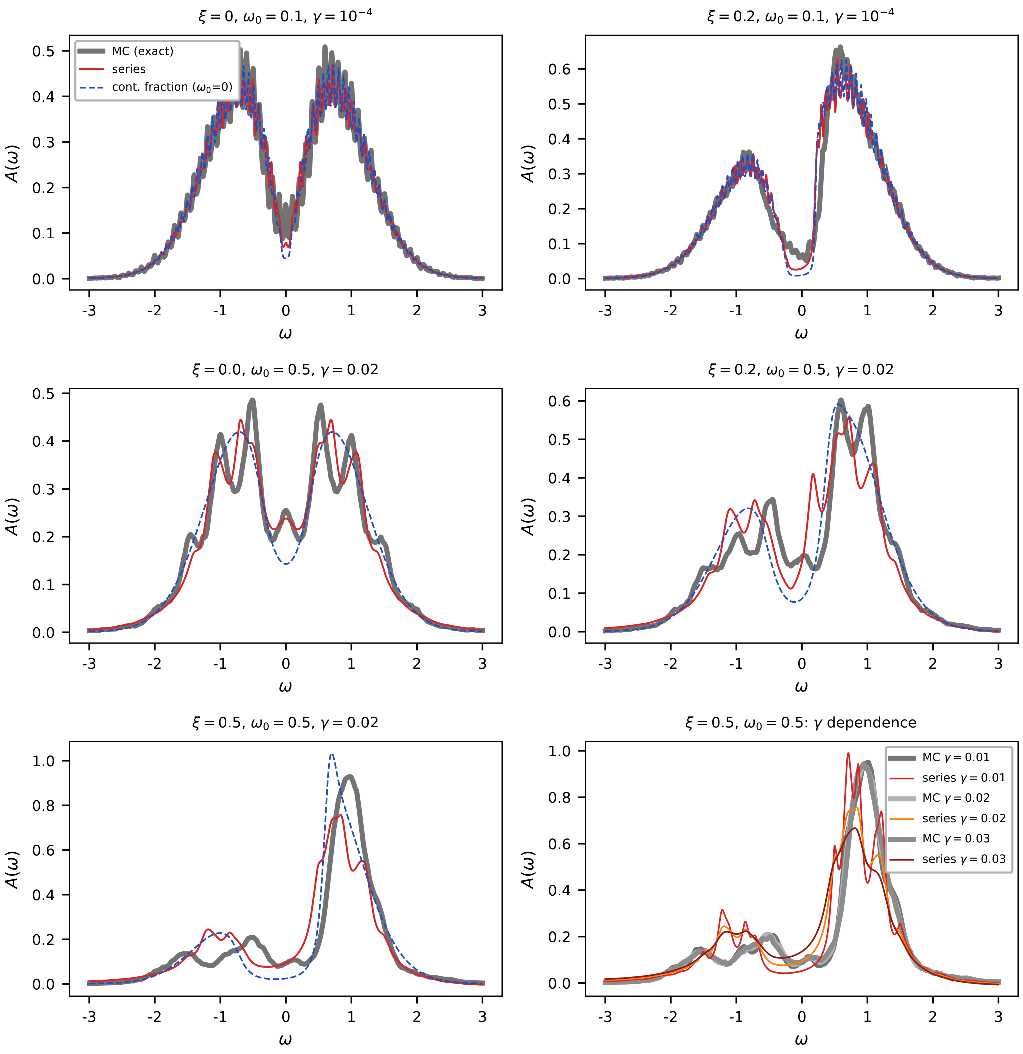}
\caption{Spectral density: incommensurate case.}
\label{fig:pp3}
\end{figure}

\begin{figure}[H]\centering\includegraphics[width=0.70\textwidth]
{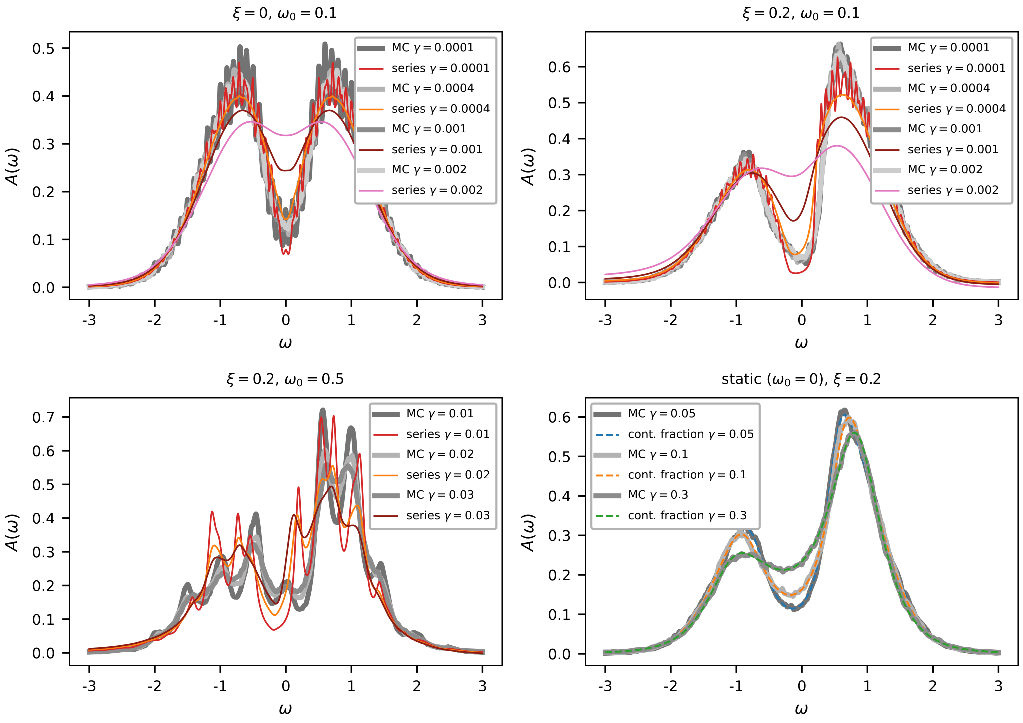}
\caption{Spectral density: incommensurate case, $\gamma$ dependence and 
static check.}
\label{fig:pp4}
\end{figure}

\begin{figure*}[t]
\centering
\includegraphics[width=0.65\textwidth]
{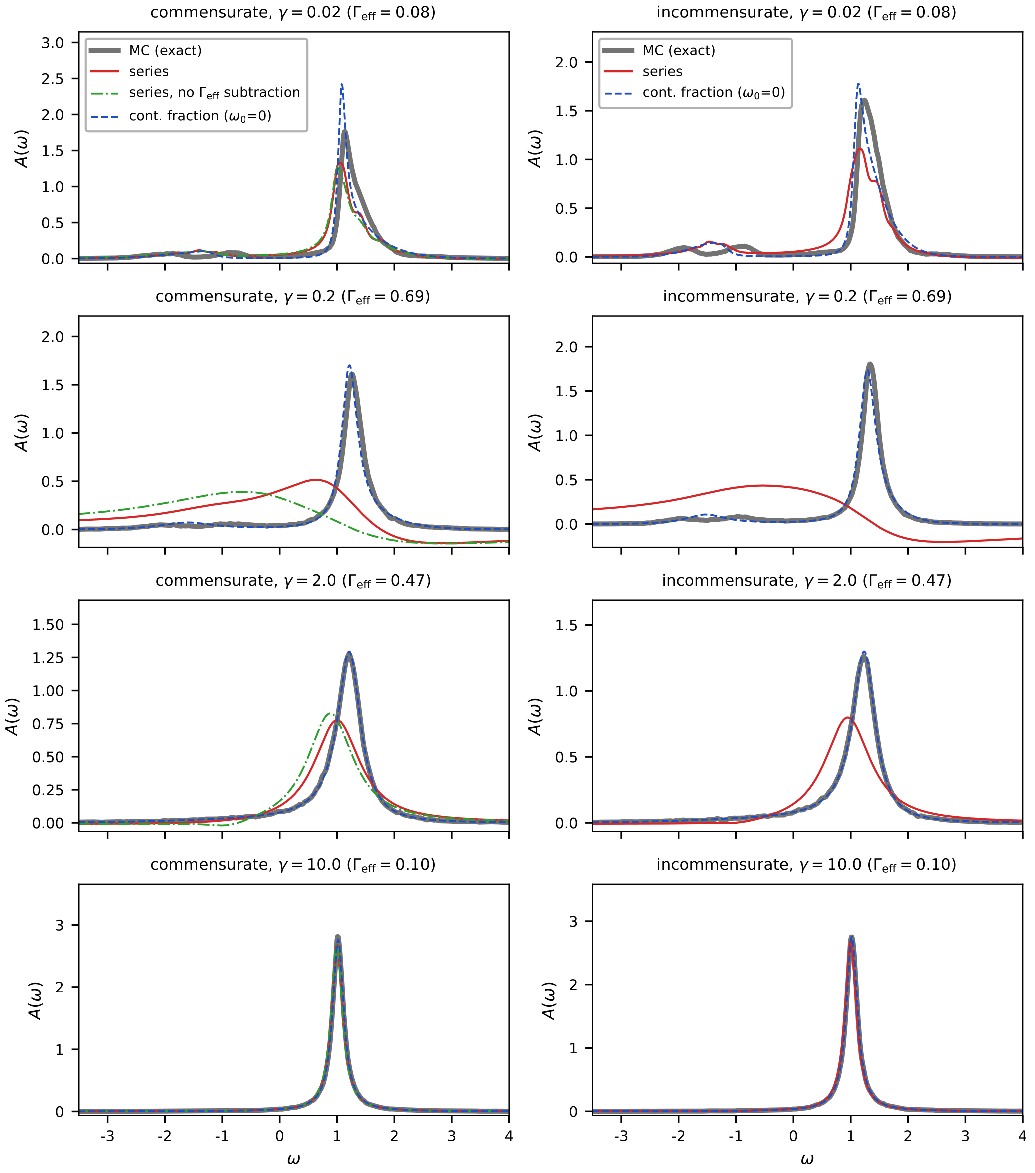}
\caption{Spectral density at $\xi=1$, $\omega_0=0.5$ for commensurate (left)
and incommensurate (right) fluctuations at $\gamma=0.02,2,10$
(top to bottom).
Grey: MC (exact); red: double series; blue dashed: continued fraction at
$\omega_0=0$ with the same $\gamma$.}
\label{fig:spectra}
\end{figure*}

\end{widetext}

\subsection{Born self-energy and the double series}

In second order in perturbation theory the electron self-energy is
\begin{equation}
\Sigma^{(2)}(\omega,\xi)=\frac{\Delta^2}{2}\left[\frac{1}
{\omega-\alpha+\xi}+\frac{1}{\omega-\beta+\xi}\right],
\label{app_Born}
\end{equation}
with $\alpha=\omega_0-i\gamma$ and $\beta=-\omega_0-i\gamma$.
To this order $G(\omega)=[\omega-\xi-\Sigma^{(2)}]^{-1}$ has, besides the main 
pole at $\omega\simeq\xi$, satellites at $\omega=-\xi\pm\omega_0$ with weights
$\Delta^2/[2(\omega_0\mp2\xi)^2]$, and the perturbation expansion becomes
singular at the resonances $2\xi=\pm\omega_0$.
At $\xi=0$ the series (\ref{BogolFreqDoubleSer}) with $u^2=v^2=\frac12$ and
to first order in $\lambda_{1,2}\propto\Delta^2$, 
the sum of the histories $(0,0)$, $(1,0)$ and $(0,1)$ reproduces the Born
result exactly in order $\Delta^2$ at $\xi=0$. It is important that the term
$-i\Gamma_{\rm eff}$ contribution to the $(0,0)$ history 
is needed for this and the identity does not hold for the series with
$\Gamma_{\rm eff}$ in the gap only.
At $\xi\neq0$ the histories $(1,0)$ and $(0,1)$ yield satellites at the
Bogoliubov energies $\pm E_{10}$,
$E_{10}=\sqrt{\xi^2+(\omega_0-i\gamma)^2}\to\sqrt{\xi^2+\omega_0^2}$,
whereas Eq.~(\ref{app_Born}) gives $\omega=-\xi\pm\omega_0$.
The two agree only for $\xi=0$; for $\xi\ll\omega_0$ they differ by
$\xi(1+O(\xi/\omega_0))$.
Third panel of Fig.~\ref{fig:bench} tests the regime
$\Delta\ll\omega_0$, where $\Gamma_{\rm eff}\ll\omega_0$. 
and hence satellites at $\omega=-\xi\pm\omega_0$ (plus shifts).
The MC satellite sits at $\omega\simeq0.78$, matching perturbation theory
($0.79$). The double series has no satellite there and instead places weight 
near 
$\sqrt{\xi^2+\omega_0^2}\simeq\pm\left(\omega_0
+\frac{\xi^2}{2\omega_0^2}\right)\simeq1.04$,
if we consider only the case of $\xi^2/\omega_0^2\ll 1$.
Its main peak is also too high ($5.35$ versus $3.64$ in MC). 
Thus the simple criterion of $\omega_0\gg\Gamma_{\rm eff}$ is not sufficient 
for the series to be reliable when $\xi\neq0$ and misplaces the sideband 
positions. At $\xi=0$ sidebands positions are predicted correctly. 
These problems are not unexpected as the effects of neglected Magnus terms 
grow with $\xi$ \cite{KS2026}.

\subsection{Spectra across the crossover}

Figure~\ref{fig:spectra} shows spectra at $\xi=1$ for representative $\gamma$
spanning the coherent side ($\gamma=0.02$, $\Gamma_{\rm eff}=0.08$),
the crossover ($\gamma=0.2$, $\Gamma_{\rm eff}=0.69$;
$\gamma=2$, $\Gamma_{\rm eff}=0.47$) and the motional-narrowing side
($\gamma=10$, $\Gamma_{\rm eff}=0.10$).
The exact spectrum is a single quasiparticle-like peak near $\omega\approx\xi$
whose properties vary smoothly with $\gamma$.
The continued fraction reproduces it increasingly well as $\gamma$ grows;
at small $\gamma$ it misses the sidebands but gives the envelope.
The double series shows three behaviours:

\begin{itemize}[nosep,leftmargin=*]

\item \emph{Coherent side} ($\gamma=0.02$): peak position within $0.05$--$0.1$,
width correct to $\sim5\%$ (commensurate) or $\sim45\%$ (incommensurate),
height $25$--$30\%$ low.

\item \emph{Crossover} ($0.1\lesssim\gamma\lesssim2$): the peak is 2--6 times
too wide, 2--4 times too low, and shifted by up to $\sim\xi$
(e.g.\ at $\gamma=0.5$, $\xi=1$ commensurate it is centred at
$\omega\approx0$ rather than $1.30$).
For commensurate fluctuations the spectral density can even become negative.
So the double series completely fails in this region.

\item \emph{Motional-narrowing side:} agreement to a few percent in position,
height and width for $\gamma\ge10$, and within $\sim10\%$ at $\gamma=5$ for
$\xi\lesssim\Delta$. 

\end{itemize}

\subsection{Density of states}
\label{sec:MCDOS}

\begin{figure*}
\includegraphics[clip=true,width=0.75\textwidth]{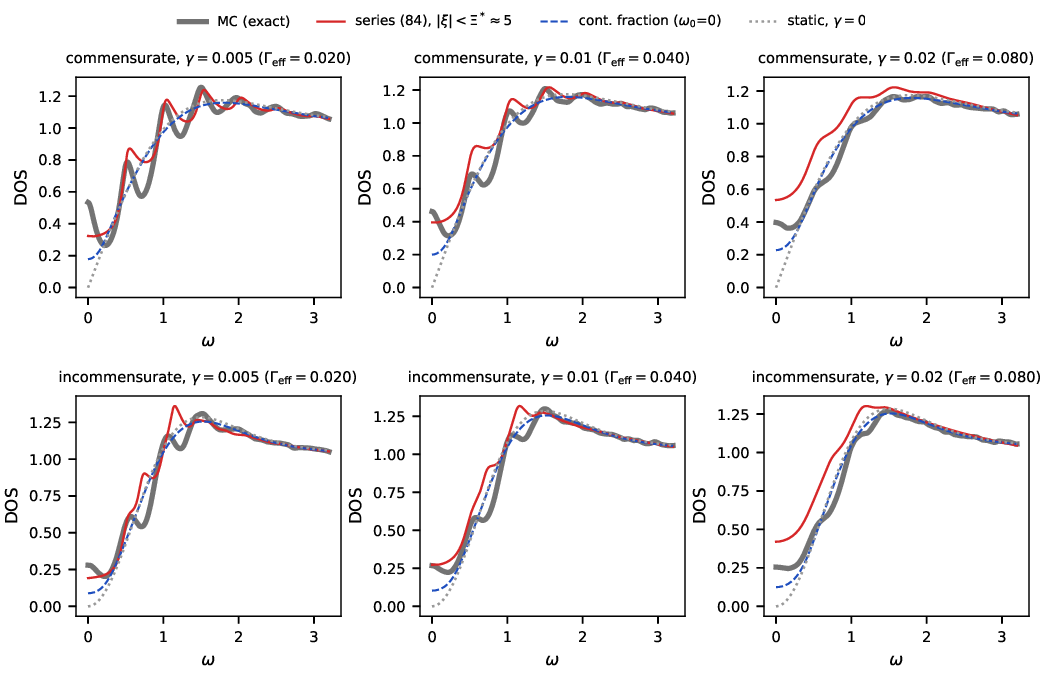}
\caption{Density of states (units of the free-electron DOS) for $\omega_0=0.5$,
$\gamma=0.005,\,0.01,\,0.02$, commensurate (upper row) and incommensurate
(lower row) fluctuations: MC (grey), series with cutoff
$\approx5\Delta$ (red), continued fraction at $\omega_0=0$ with the same
$\gamma$ (blue dashed), and static result at $\omega_0=\gamma=0$ (dotted). $\eta=0.04$.}
\label{MC_dos}
\end{figure*}

The density of states  was obtained from the MC solution by
integrating $A(\omega,\xi)$ over $\xi$ with $\eta=0.04$.
Figure~\ref{MC_dos} compares it with the series, the continued fraction and the 
static results for $\omega_0=0.5$ and $\gamma=0.005,\,0.01,\,0.02$. 
The exact DOS has a
pseudogap and additional peaks at multiples of 
$\omega_0$ ($\omega\simeq0.5,\,1,\,1.5$),
which are suppressed with growing $\gamma$.
The series reproduces the positions of the modulation peaks at
$\gamma=0.005$ and $0.01$, but the contrast of the modulation is too small
(e.g.\ for commensurate fluctuations, $\gamma=0.005$, the first minimum is
$0.79$ against $0.57$ in MC), and at $\gamma=0.02$ the pseudogap region
is somewhat too filled.

The integral over $\xi$ of the series spectral function is not absolutely
convergent, 
so that the DOS depends logarithmically
on the cutoff  of the integration. 
In the calculations of Fig.~\ref{MC_dos} the cutoff
was therefore fixed by requiring that the series DOS coincides with the
continued-fraction DOS at large $\omega$ ($\omega=3$);
this gives the cutoff $\approx5\Delta$ ($4.9$--$5.1$) in all cases.

\section{Conclusion}

An exact MC solution of the dynamic finite-$Q$ pseudogap model
confirms the exact limits of the double series of \cite{KS2026} and the
exactness of the static continued fraction, and shows that the coherent double
series is a good description for $\Gamma_{\rm eff}\lesssim0.15\,\omega_0$ and 
in the motional-narrowing limit, while failing qualitatively in the crossover 
region. More precisely, our MC data show that the series is useful for 
$\Gamma_{\rm eff}\lesssim0.15\,\omega_0$ and for
$\gamma\gtrsim10\,\omega_0$ ($\gamma\ge5$ here); in between, the static
continued fraction is the more accurate description once
$\Gamma_{\rm eff}\gtrsim\omega_0$, and already at 
$\gamma\gtrsim0.2$ ($\Gamma_{\rm eff}\approx0.7$,
$\gamma\approx0.4\,\omega_0$). 
The crossover is smooth in the exact solution.
The Monte Carlo benchmark can serve as a test-bed
for improved approximations, for example including the leading Magnus
corrections \cite{Magnus}. 
The future efforts here may probably be undertaken
along the lines developed in diagram techniques for spin-operators
\cite{VLP,IK,IS,BKY,BL}.
Overall, the general qualitative picture of sideband formation is exact 
for $\xi=0$ in commensurate case and quite satisfactory  quantitatively for
incommensurate case. This is very important as most ARPES experiment are
usually done for $\xi=0$, i.e. at the Fermi surface, or close to it. 
Finite $\xi$ case double series provides rather satisfactory description 
at small $\gamma$, when $\Gamma_{\rm eff}$ is small, as compared to $\omega_0$,
and for small enough values of $\xi$, giving qualitatively reasonable 
picture\footnote{It is even quantitative for $|\xi|<0.3\gamma$ (Supplement S2)},
as compared with exact results. The double series fails qualitatively 
in crossover region, but the continuous fraction description is already very 
good there. These results essentially confirm the general physical picture of dynamic
pseudogap formation presented in Ref. \cite{KS2026}, though with some 
limitations.

\subsection{Acknowledgements}

The author is grateful to E.Z. Kuchinskii for useful discussions and
help during this work.

\clearpage

\appendix

\begin{widetext}

\section*{Supplemental Material}

\setcounter{section}{0}
\renewcommand{\thesection}{S\arabic{section}}
\renewcommand{\thefigure}{S\arabic{figure}}
\renewcommand{\thetable}{S\arabic{table}}
\setcounter{figure}{0}\setcounter{table}{0}

\section{Monte Carlo method details and some additional results}

\paragraph{Field generation.} A complex OU process $z$ with decay constant
$\kappa$ and stationary variance $s$ is advanced by the exact transition
$z_{j+1}=a\,z_j+\sqrt{s(1-|a|^2)}\,\zeta_j$, $a=e^{-\kappa\,\mathrm dt}$,
with $\zeta_j$ unit-variance circular complex Gaussian and $z_0$ drawn from
the stationary distribution; then $\avg{z(t)z^*(0)}=s\,e^{-\kappa|t|}$.
For commensurate case
$\Delta_1=\mathrm{Re}\,z$, $\kappa=\gamma+i\omega_0$, $s=2\Delta^2$, so that
$\avg{\Delta_1(t)\Delta_1(0)}=\Delta^2e^{-\gamma|t|}\cos\omega_0t$.
For incommensurate case $\Delta=a+b$ with independent $a,b$,
$\kappa=\gamma\pm i\omega_0$, $s=\Delta^2/2$ each, giving  Eq.~(\ref{eq:corr}). 
For $\omega_0=0$ this reduces to a single complex
OU process of variance $\Delta^2$.

\paragraph{Propagation.} On each step the field is replaced by the average
$\bar\Delta_j$ of its endpoint values and the step operator is the exact
exponential
$\hat M_j=\cos(E_j\mathrm dt){\hat 1}-i\,\sin(E_j\mathrm dt)\,\hat H_j/E_j$,
$E_j=\sqrt{\xi^2+|\bar\Delta_j|^2}$,
$\hat H_j=\begin{pmatrix}\xi&\bar\Delta_j^*\\\bar\Delta_j&-\xi\end{pmatrix}$
(real $\bar\Delta_j$ for commensurate fluctuations), using
$\hat H^2=E^2{\hat 1}$. The only discretization error is from the step-average
of the field. The sample mean of $U_{11}$ is recorded at every step.
The spectrum is the trapezoid-rule evaluation of the transformation
on a uniform frequency grid; $e^{-\eta t}$ both windows
the finite $T$ (neglected tail $e^{-\eta T}$) and broadens every
feature by $\eta$. The same $\omega\to\omega+i\eta$ was used in the series
and continued fraction.

\begin{table}[H]
\centering
\caption{Simulation settings ($\Delta=1$). $N$ is the number of realizations
per curve; in all runs $\eta=0.02$ unless noted.}
\begin{tabular}{lccccc}
\toprule
Set & $\gamma$ & $T$ & $\mathrm dt$ & $\eta$ & $N$ \\
\midrule
$\omega_0=0.1$ & $10^{-4}$--$2\times10^{-3}$ & 600 & 0.05 & 0.01 & 30000 \\
$\omega_0=0.5$ & 0.005--0.03 & 300 & 0.04 & 0.02 & 30000 \\
Static check, $\omega_0=0$ & 0.05--0.3 & 200 & 0.04 & 0.02 & 30000 \\
Scan, $\omega_0=0.5$ & 0.02, 0.05, 0.1 & 300 & 0.04 & 0.02 & 20000 \\
Scan, $\omega_0=0.5$ & 0.2, 0.5, 1, 2 & 200 & 0.04 & 0.02 & 20000 \\
Scan, $\omega_0=0.5$ & 5 / 10 / 20 & 200 / 250 / 300 & 0.02 / 0.01 / 0.005 & 0.02 & 15000 \\
Weak coupling & 0.05 ($\Delta=0.3$, $\omega_0=1$) & 120 & 0.02 & 0.05 & 20000 \\
\bottomrule
\end{tabular}
\end{table}

\paragraph{Time step.} At $\gamma=10$ (steps $0.02,0.01,0.005$) and
$\gamma=20$ ($0.01,0.005,0.0025$), $\xi=1$, commensurate, $N=8000$, $T=60$,
$\eta=0.05$, the maximum differences between runs were $0.034$--$0.046$ with no
systematic decrease with $\mathrm dt$, i.e.\ noise-dominated, so no time-step
bias is visible at this level. Production runs use $\gamma\,\mathrm dt=0.1$ at
large $\gamma$ and $\mathrm dt\le0.05$ otherwise. We did not perform a
systematic $\mathrm dt$ study for $\gamma<10$.

\paragraph{Statistical noise.} A two-seed comparison ($N=2\times10^4$,
$\eta=0.05$, $\xi=0.5$, $\omega_0=0.5$, $\gamma=0.02$) gave a maximum-norm
error $0.015$ in $A(\omega)$. Scaling roughly as $\eta^{-1/2}N^{-1/2}$
suggests $0.02$--$0.04$ for the production runs; this is an extrapolated
estimate, not measured separately.

\begin{figure}[t]
\centering
\includegraphics[width=0.75\textwidth]{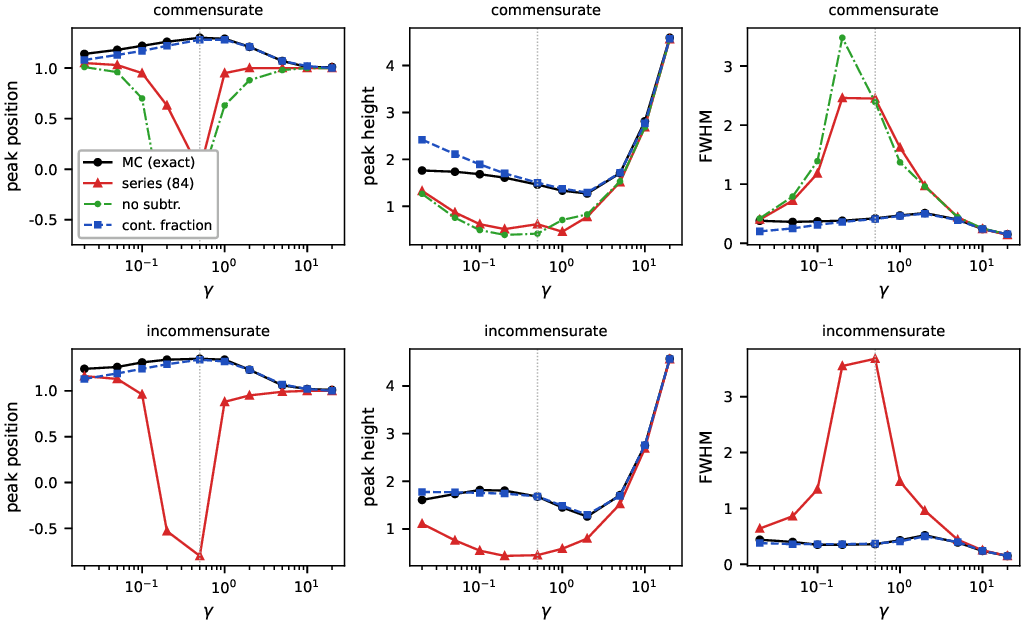}
\caption{Main-peak position, height and FWHM at $\xi=1$ versus
$\gamma$ ($\omega_0=0.5$, dotted line: $\gamma=\omega_0$) for MC (black),
series (red) and continued fraction (blue).}
\label{fig:peak}
\end{figure}

\section{Peak properties and Domain of Validity}

Figure~\ref{fig:peak} tracks the position, height and FWHM of the main peak
at $\xi=1$ against $\gamma$ for MC, series and continued fraction. The exact
peak parameters change smoothly and monotonically over three decades of
$\gamma$, and the continued fraction follows them closely for $\gamma\gtrsim0.2$.
The series deviates strongly in a window around $\gamma\sim\omega_0$: its
position moves to $0$ (commensurate) or negative $\omega$ (incommensurate),
its height drops by a factor 3--4 and its width exceeds $2$. It converges to
the exact result for $\gamma\gtrsim5$ ($\gtrsim10\,\omega_0$).

The numerical data allow us to formulate the domain of applicability of the
series more precisely.  We have compared the main-peak position, height and
width in the MC and series spectra for commensurate
fluctuations in a set of 81 parameter points with $\omega_0/\Delta=0.25$--$2$,
$\gamma/\Delta=0.01$--$20$ and $|\xi|/\Delta=0.1$--$2$, and called the 
agreement ``qualitative'' if the position agrees within half of the exact 
linewidth and the height and width within 30\%. 
We find that all 26 points with
\begin{equation}
|\xi|\lesssim\frac{\gamma}{3}
\label{ValidXiGamma}
\end{equation}
show agreement, with no exceptions, for all values of $\omega_0$
(Fig.~\ref{MC_valid}); disagreement starts at $|\xi|/\gamma\approx0.35$--$0.5$.
A few cases with larger $|\xi|/\gamma$ also agree, when $|\xi|\lesssim0.1\Delta$
(close to the nested Fermi surface, where the series is exact) or marginally.
The criterion $\omega_0\gg\Gamma_{\rm eff}$, on the other hand, does not separate the cases
well: of 37 points with $\Gamma_{\rm eff}/\omega_0\le0.3$ only 18 agree.
The condition (\ref{ValidXiGamma}) has a simple interpretation.
The non-Abelian contribution comes from the commutator 
proportional to $\xi[\Delta(t_2)-\Delta(t_1)]$. It is suppressed when the field
decorrelates within a time $\gamma^{-1}$ short compared with
the time $\xi^{-1}$ in which the phase difference between the two nested
branches builds up, i.e.\ for $\xi\gamma^{-1}\lesssim1/3$
(the motional-narrowing condition for the non-commuting part).
In the coherent regime $\gamma\to0$ the condition cannot be satisfied except
for $\xi\to0$, which explains why the series is accurate at the nested Fermi
surface and displaces the sidebands elsewhere. The threshold in
(\ref{ValidXiGamma}) is empirical, was obtained for commensurate fluctuations
only, and depends on our choice of the agreement criteria; we did not attempt
to establish an analogous condition for incommensurate fluctuations,
for which the series is not exact even at $\xi=0$.

\begin{figure}[t]
\centering
\includegraphics[width=0.75\textwidth]{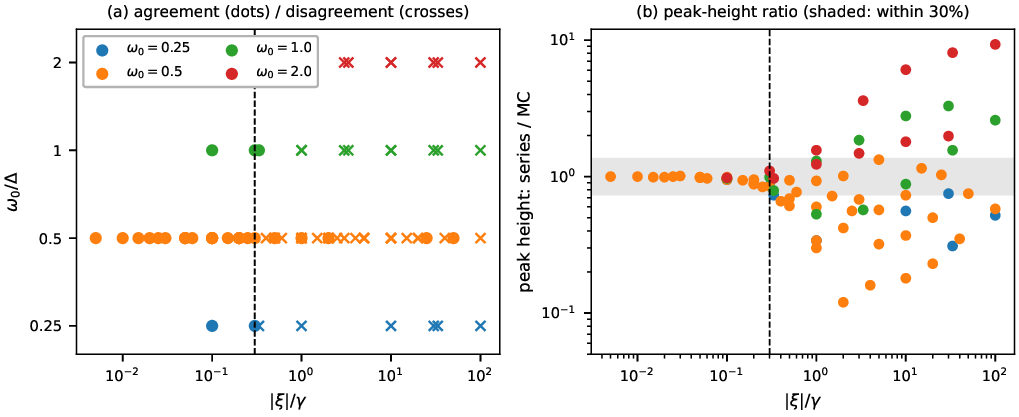}
\caption{Domain of agreement between series and MC 
(commensurate fluctuations, 81 parameter points). 
(a) Agreement (dots) and disagreement (crosses) in the plane
$(|\xi|/\gamma,\ \omega_0/\Delta)$; the dashed line 
is $|\xi|/\gamma=0.3$. 
(b) Ratio of peak heights (series/MC) versus $|\xi|/\gamma$; 
shaded: within 30\%. Different colours correspond to different $\omega_0$.}
\label{MC_valid}
\label{fig:err}
\end{figure}

\end{widetext}


\begin{thebibliography}{99}



\bibitem{Sadovskii1974} M.V.~Sadovskii, \emph{A Model of a Disordered System},
Sov.\ Phys.\ JETP \textbf{39}, 845
(1974).

\bibitem{Sadovskii1979} M.V.~Sadovskii, \emph{Exact Solution for the Density of
Electronic States in a Model of a Disordered System}, Sov.\ Phys.\
JETP \textbf{50}, 989 (1979).

\bibitem{Sadovskii2019} M.V.~Sadovskii, \emph{Diagrammatics},
2nd ed.\ (World Scientific, 2019).

\bibitem{KS2026} E.Z.~Kuchinskii and M.V.~Sadovskii,
\emph{Dynamic Pseudogap Model}, arXiv:2608.19592v2.

\bibitem{KiselevKikoin2009} M.N.~Kiselev and K.~Kikoin,
\emph{Scalar and Vector Keldysh models in the time domain},
JETP Lett.\ \textbf{89}, 133 (2009).

\bibitem{KS2024} E.Z.~Kuchinskii and M.V.~Sadovskii,
\emph{Generalized Dynamical Keldysh Model}, JETP \textbf{166}, 42 (2024);
arXiv:2401.08118.

\bibitem{Magnus} W.~Magnus, \emph{On the exponential solution of differential
equations for a linear operator}, Commun.\ Pure Appl.\ Math.\ \textbf{7},
649 (1954).

\bibitem{Kubo} R.~Kubo, J.\ Phys.\ Soc.\ Jpn.\ \emph{Note on the Stochastic 
Theory of Resonance Absorption},  {\bf 9}, 935 (1954).

\bibitem{VLP}V.G. Vaks, A.I. Larkin. S.A. Pikin. {\it Spin Waves and Correlation
Functions in a Ferromagnet}. JETP {\bf 26}, 647 (1968); [Zh. Eksp. Teor. Fiz.
{\bf 53}, 1089 (1968)].

\bibitem{IK}Yu.A. Izyumov, F.A. Kassan-Ogly, Yu.N. Skryabin.
{\it Diagram Technique for Spin-Operators and its Applications to Some Problems
od Ferromagnetism}. J. de Physique {\bf 32}, C1-86 (1971)

\bibitem{IS}Yu.A. Izyumov, Yu.A. Skryabin. {\it Statistical Mechanics of
Magnetically Ordered Systems} (Consultants Bureau, NY, 1988)

\bibitem{BKY}V.G. Bary'akhtar, V.N. Krivoruchko, D.A, Yablonskii,
{\it Spin Green functions and the problem of summation over physical states}.
JETP {\bf 58}, 351 (1983); [Zh. Eksp. Teor. Fiz. {\bf 88}, 601 (1983)]

\bibitem{BL}V.I. Belinicher, V.S. L'vov. {\it Spin diagram technique for
non-equilibrium processes in the theory of magnetism}. JETP {\bf 59}, 564
(1984); [Zh. Eksp. Teor. Fiz. {\bf 86}, 967 (1984)].


\end{thebibliography}
\end{document}